\documentclass{ws-ijmpc}
\usepackage{epsfig,graphics}

\begin{document}

\catchline{}{}{}{}{}
%%%%%%%%%%%%%%%%%%%%%%%%%%%%%%%%%%%%%%%%%%%%%%%%%%%%%%%%%%%%%%%%%%%%%

\title{RURAL-URBAN MIGRATION IN D-DIMENSIONAL LATTICES}

\author{AQUINO L. ESP\'INDOLA}
\address{Instituto de F\'{\i}sica, Universidade Federal Fluminense,\\
Niter\'oi, RJ 24.210-340,
Brazil\\
aquino@if.uff.br}

\author{T. J. P. PENNA}
\address{Instituto de F\'{\i}sica, Universidade Federal Fluminense,\\
Niter\'oi, RJ 24.210-340,
Brazil\\
tjpp@if.uff.br}

\author{JAYLSON J. SILVEIRA}
\address{Depto de Economia, Universidade Estadual Paulista - UNESP,\\
Araraquara, SP,
Brazil\\
jaylson@fclar.unesp.br}

\maketitle

\begin{history}
\received{06 Jun 2005}
\revised{Day Month Year}
\end{history}

\begin{abstract}
The rural-urban migration phenomenon is analyzed by using an agent-based computational model. Agents are placed on lattices which dimensions varying from $d=2$ up to $d=7$. The localization of the agents in the lattice define their social neighborhood (rural or urban) not being related to their spatial distribution. The effect of the dimension of lattice is studied by analyzing the variation of the main parameters that characterizes the migratory process. The dynamics displays strong effects even for around one million of sites, in higher dimensions ($d=6$, $7$).

\end{abstract}

\ccode{PACS Nos.: 89.65.Gh, 05.10.-a, 82.20.Wt, 05.50.+q}

\section{Introduction}
\label{intriduction}

The rural-urban migration is  a very important phenomenon which occurs in developing economies. In our previous work, see Ref. \refcite{nos}, we analyzed such a phenomenon using an agent-based computational model, considering the rural-urban migration as a discrete choice problem with social interaction.\cite{brock} The migratory decision was modelled taking into account the pressure caused by the different earnings between rural and urban sectors and the neighborhood influence.  The influence caused by neighbors was modelled like in the Ising model in a two-dimensional lattice in the presence of an external field.

\par Simulations in this model show some emergent properties which are consistent with historical data of developing countries, namely: a transitional dynamics characterized by increasing of population fraction living at urban area and {\it per capita} income, followed by equalization of expected wages between rural and urban sectors (Harris-Todaro equilibrium condition) and urban concentration. In this paper we extend this analysis to study the behavior toward equilibrium and its properties in $d$-dimensional lattices, $3 \le d \le7$, instead of $d=2$ in Ref. \refcite{nos}.

\section{The benchmark model}
\label{benchmakmodel}

In this section we present the general equilibrium structure of a two-sector less developed economy set in Ref. \refcite{nos} to study the rural-urban migration process.\cite{ray,willianson} We also review the formalization of the social interaction modelled by an Ising like model to as proposed in Ref. \refcite{nos}. These results will be taken to carry out the generalizations of subsection \ref{transitionalnd}.

\subsection{The general equilibrium structure of a rural-urban economy}
\label{generalequilibrium}

There are two productive sectors considered: urban and rural. The urban sector is formed by firms which are specialized in the production of manufacturated goods whereas the rural sector is formed by farms which produce agricultural goods.

The production of the manufacturing sector is given by\cite{day}

\begin{equation}
Y_m = \xi_1 N_u^\alpha,
\label{ym}
\end{equation}
where $N_u$ is the amount of workers of the urban sector. $\xi_1>0$ and $\alpha>0$ are parametric constants.

The equilibrium wage of this sector is
\begin{equation}
w_m=\xi_2 N_u^{\alpha-1},
\label{wm}
\end{equation}
where $\xi_2>0$ is a parametric constant.

The aggregated production of the rural sector  is given by 
\begin{equation}
Y_a=\xi_3(N-N_u)^\phi,
\label{ya}
\end{equation}
where $N$ is the total number of workers of the economic system. $\xi_3>0$ and  $\phi>0$ are parametric constants.

The equilibrium wage is
\begin{equation}
w_a= \xi_4 p (N-N_u)^{\phi-1},
\label{wa}
\end{equation}
where $\xi_4>0$ is a parametric constant.

The terms of trade between these sectors \cite{harristodaro,todaro} is measured by the price $p$:

\begin{equation}
p=\rho\left(\frac{Y_m}{Y_a}\right)^\gamma,
\label{p}
\end{equation}
 where $\rho>0$ and $\gamma>0$ are parametric constants.

 Given the equilibrium urban unemployment rate, $u$, as determined in Ref. \refcite{nos}, using  Eqs. (\ref{ym}) and (\ref{wm}) one can calculate the state of the urban sector. In a similar manner, the rural sector has its state calculated by using Eqs.  (\ref{ya}), (\ref{wa}) and (\ref{p}). The complete deduction of the equations of state of both sectors are done in Ref. \refcite{nos}. Typical values for parameters mentioned above are $u=0.10$, $\xi_1=144.75$, $\xi_2=112.59$, $\xi_3=500.00$, $\xi_4=150.00$, $\alpha=0.70$, $\phi=0.30$, $\rho=1.00$, $\gamma=1.25$ and $\beta=3.00$.

\subsection{The sectorial migration model}
\label{sectorial migration}

The migration process was modelled by an agent-based computational model. The decision of migrating or not is considered taking into account the difference of expected wages between the sectors, called deterministic private utility, and the influence that individuals suffer by the group they are included in, called deterministic social utility.\cite{brock,durlauf,gustavo} As mentioned before, only discrete choices are allowed, so each worker has it state defined by $\sigma_i \in \left\{-1,+1\right\}$, where $\sigma_i=-1$ represents a rural worker and $\sigma_i=+1$ represents an urban worker.
 
As mentioned above, during the decision process each worker takes into account explicit and observable incentives and the influence of their social neighborhood. Then, the total (private and social) utility is given by

\begin{equation}
{\mathcal{H}}_i=K\left[(1-u)w_m-w_a\right]\sigma_i + J\sum_{j\in n_i}\sigma_i\sigma_j,
\label{hi}
\end{equation}
where $K>0$ and $J>0$ are parametric constants.

The probability that each workers reviews his/her sectorial decision is given by the activity $a$ defined in Ref. \refcite{thadeu}. This parameter guarantees that only a fraction of the total population review their decision becoming potential migrants. The probability that a potential migrant becomes an actual migrant is given by a cumulative distribution:

\begin{equation}
Pr_i=\frac{1}{1+e^{-\beta{\mathcal{H}}_i}},
\label{pri}
\end{equation}
where $\beta>0$ is a parametric constant which measures the heterogeneity of agents.

The probability that a worker $i$ migrates or not depends on the probability calculated in Eq. (\ref{pri}). Then, the higher the value of the total utility, Eq. (\ref{hi}), the higher the probability that the worker does not change his/her sectorial decision.

\section{Properties of the transitional dynamics}

In our previous work we analyzed the transitional dynamics of the rural-urban migration in a two-dimensional lattice.  The determination of the macrostate of the system in each time step can be done by using Eqs. (\ref{ym}-\ref{p}). In this section we will briefly review some of these results\footnote{For further details see Ref. \refcite{nos}}.

\par
\begin{figure}[hbt]
\centerline{\psfig{file=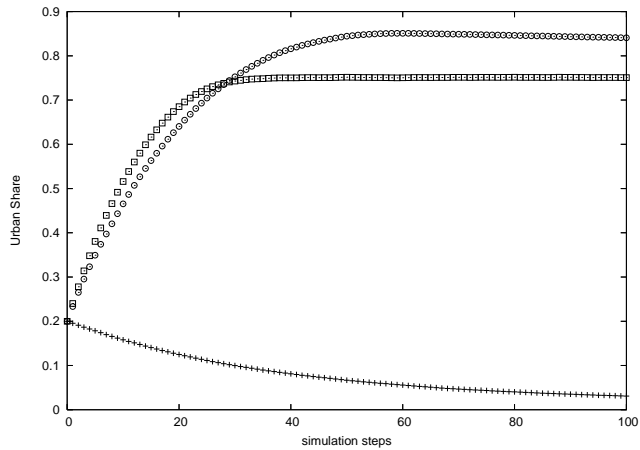,height=6.0cm,angle=0}}
%\vspace*{8pt}
\caption{Proportion of workers at urban sector as function of time for three different set of parameters $J$ and $K$. Circles: $(J>0$, $K>0)$; Squares: $(J=0$, $K>0)$; Crosses: $(J>0$, $K=0)$}
\label{nut}
\end{figure}

 The main variable that characterizes the migratory process is the fraction of workers allocated in the urban sector, $n_u$, also called urban share. In Fig. \ref{nut} one can see the urban share as function of time. The three curves of this Figure are plotted for different combination of the parameters $K$ and $J$ of Eq. (\ref{hi}). The set  $(J=0$, $K>0)$ plotted in Fig. \ref{nut} means that agents review their sectorial position taking into account only the deterministic private utility, ignoring the neighborhood interaction. The second case,  $(J>0$, $K>0)$, both of the effects mentioned before are considered in the reviewing process. In the last case,  $(J>0$, $K=0)$, only neighborhood influence in considered in the reviewing process what make this case an unrealistic one.

\subsection{Transitional dynamics in d-dimensions}
\label{transitionalnd}

The simulations for $d$-dimensional lattices were carried out placing workers in lattices of dimensions $d=3,~4,~5,~6$ and $7$. This dimensional change in the lattice will modify the number of nearest neighbors that each worker will have and this relation will depend on the dimension of the lattice:

\begin{equation}
n_b=2d;
\label{neighbors}
\end{equation}
where $n_b$ is the number of nearest neighbors and $d$ is the dimension of the lattice. In all simulations the initial urban share is $n_u=0.2$, what means that $20\%$ of the workers are located in the urban sector. All others parameters of the system are the same, if not, it will be mentioned.

\begin{figure}[hbt]
\centerline{\psfig{file=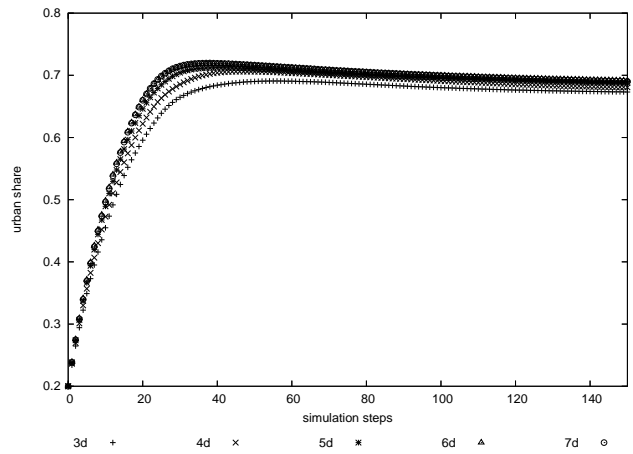,height=6.0cm,angle=0}}
%\vspace*{8pt}
\caption{Urban share as function of time for different dimensions of lattice ($J=6.0$, $K=2.0$). The linear dimension of each lattice is $L_{(3d)}=100$, $L_{(4d)}=32$, $L_{(5d)}=16$, $L_{(6d)}=10$, $L_{(7d)}=7$. An overshooting in the urban share becomes more evident when the dimension is increased.}
\label{nund}
\end{figure}

 In Fig. \ref{nund} the urban share is plotted as function of time. In this figure one can see that the increasing of the dimension of the lattice will accelerate the migratory process provoking an overshooting in the urban share as observed in developing economies. On the other hand, smaller dimension takes the system to reach equilibrium faster than higher dimension lattices. In Eq. (\ref{neighbors}) one can see that the variation in the size of the lattice will change the amount of nearest neighbors that each worker is connected. Therefore, as the whole parameters are kept constant to all simulations, the behavior of the curves in Fig. \ref{nund} are strictly related to the number of nearest neighbors in $d$-dimensional lattices.

\begin{figure}[hbt]
\centerline{\psfig{file=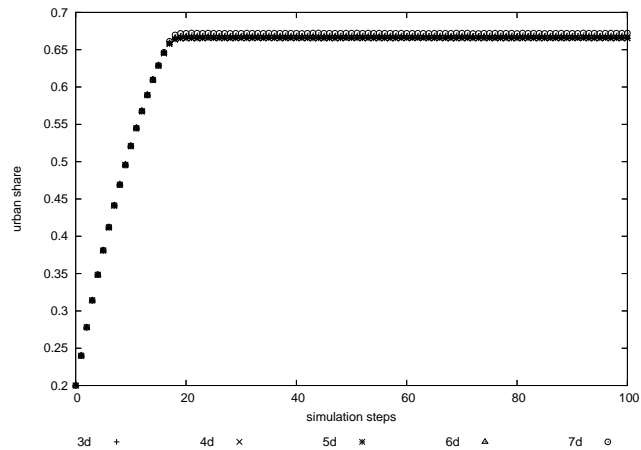,height=6.0cm,angle=0}}
%\vspace*{8pt}
\caption{Urban share as function of time for different dimensions of lattice for $J=0$. The linear dimension of each lattice is $L_{(3d)}=100$, $L_{(4d)}=32$, $L_{(5d)}=16$, $L_{(6d)}=10$, $L_{(7d)}=7$.}
\label{nundJzero}
\end{figure}

\begin{figure}[hbt]
\centerline{\psfig{file=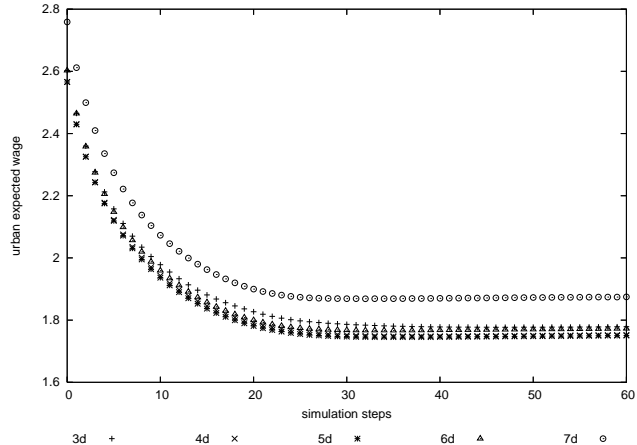,height=6.0cm,angle=0}}
%\vspace*{8pt}
\caption{Wage of the manufacturing sector $w_m$ for different dimensions of lattice. The linear dimension of each lattice is $L_{(3d)}=100$, $L_{(4d)}=32$, $L_{(5d)}=16$, $L_{(6d)}=10$, $L_{(7d)}=7$.}
\label{wmnd}
\end{figure}

The statement above is proved in Figure \ref{nundJzero} where all the simulations were carried out as in Fig. \ref{nund} but keeping parameter $J=0$. The parameter $J$, in Eq. (\ref{hi}), set null means that the interaction among neighbors is not being considered in the reviewing process done by the agents. Then, this figure clearly shows that migratory process is being modified due the number of neighbors, i.e., the social interaction is playing more important role in the decision of migrate or not.  

\par
\begin{figure}[hbt]
\centerline{\psfig{file=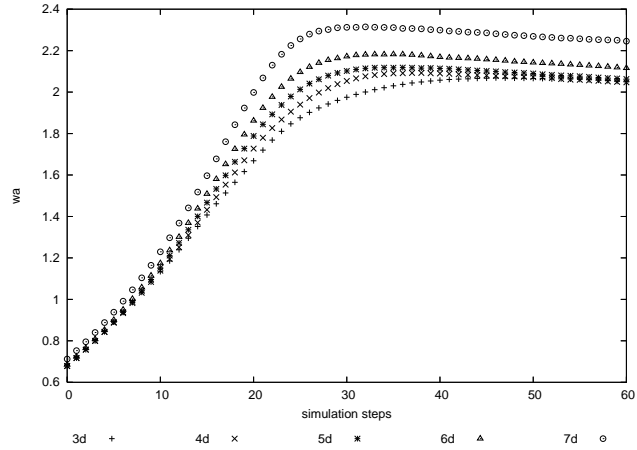,height=6.0cm,angle=0}}
%\vspace*{8pt}
\caption{Wage of rural sector $w_a$ for different dimensions of lattice. The linear dimension of each lattice is $L_{(3d)}=100$, $L_{(4d)}=32$, $L_{(5d)}=16$, $L_{(6d)}=10$, $L_{(7d)}=7$.}
\label{wand}
\end{figure}

\begin{figure}[hbt]
\centerline{\psfig{file=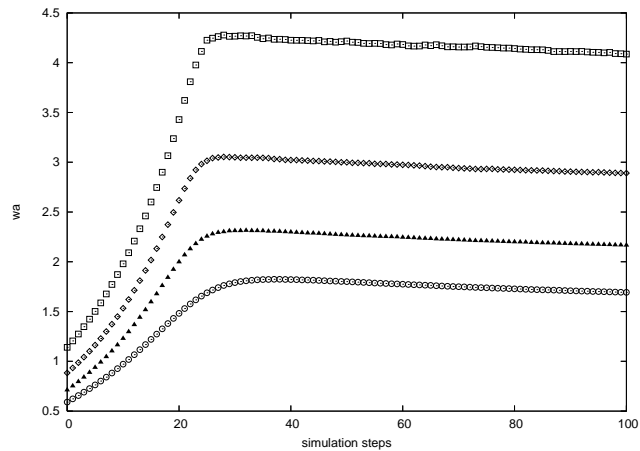,height=6.0cm,angle=0}}
%\vspace*{8pt}
\caption{Wage of rural sector $w_a$ for different linear length of lattice for 7d. Squares: $L=5$; Diamonds: $L=6$; Triangles: $L=7$; Circles: $L=8$.}
\label{wa7d}
\end{figure}

 Figures \ref{wmnd} and \ref{wand} are plotting of expected urban wage $w_m$ and the rural wage $w_a$, respectively. Once again, the only parameter which is changed is the dimension of the lattice. In these figures is shown that the curves of $w_m$ and $w_a$ have a similar behavior independently of the dimension of lattice. It is important to mention that the curves plotted for $d=7$ in Figs. \ref{wmnd} and \ref{wand} are the only ones in which the total number of workers $N$ is different from the values used in lower dimension lattices due to numerical limitations. Our sizes are considerably smaller than the world record simulated in Ref. \refcite{stauffer}, because we do not use multi-spin coding and $w_a$ and $w_m$ must be reevaluated after each step. Therefore, in Fig. \ref{wa7d} we plotted the rural wage as function of time for lattice of $d=7$, varying $N$ in order to check its influence in the results.

\par The variable $r_e\equiv(1-u)w_m/w_a$ measures the expected wages ratio. Observing Figures \ref{wmnd} and \ref{wand} one can see that $r_e\approx1.0$, what indicates that the expected urban wage and the rural wage converge to same value. This is the Harris-Todaro equilibrium condition.\cite{harristodaro}

\begin{figure}[hbt]
\centerline{\psfig{file=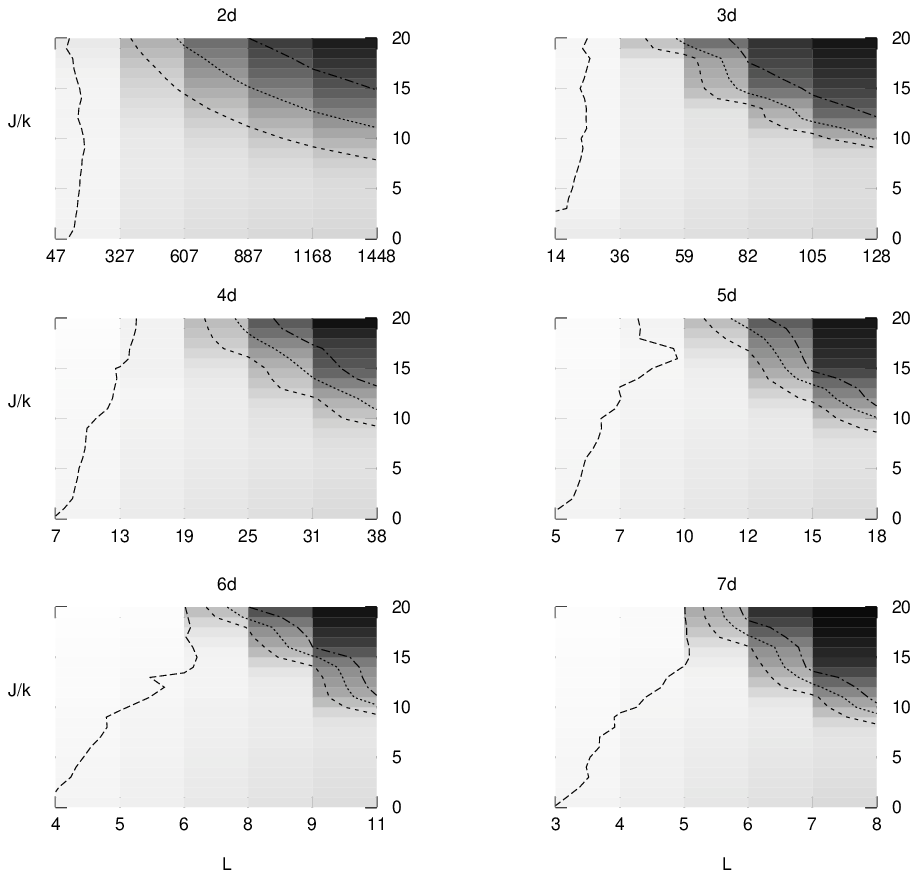,height=14.0cm,angle=0}}
%\vspace*{8pt}
\caption{Urban share $n_u$ as function of the ratio $J/K$ and the lattice size $L$. Lighter areas correspond to higher urban share and darker areas to lower urban share.}
\label{JkL}
\end{figure}

\par The deterministic total utility of each worker is calculated by using Eq. (\ref{hi}), where can see two differrent interactions acting in the worker's decision. The first part of right side of Eq. (\ref{hi}) acts like an external time dependent field and in this case is related to the difference of wages between rural and urban sector. The intensity of this field is calibrated by the constant $K$. The second part of right of Eq. (\ref{hi}) is the interaction of a worker among his/her neighborhood and it is related to the influence caused by the neighbors which intensity depends on the value of the constant $J$.

\par Therefore, in Ref. \refcite{nos}, to study the influence of the variation of these parameters we ran several simulations where the ratio $J/K$ and the size of the $L$ so that we could check their effects in the values of the urban share. Now, we apply the same procedure to lattices with dimensions $d>2$.

\par Figure \ref{JkL} has plotted the urban share $n_u$ as function of ratio $J/K$ for different lattice size $L$.  Each figure of this set is done for different dimension from $d=2$, top left, to $d=7$, right bottom . The horizontal axis of each figure is the linear dimension of the lattice $L$, then the number of workers is obtained by the relation $N=L^d$. 

\par All sets of Fig. \ref{JkL} reveal a similar distribution of the equilibrium values of the urban share, $n_u$, for the six different dimensions simulated. The equilibrium urban share is slightly different when the spatial dimension $d$ is changed. This result is in agreement with Fig. \ref{nund} where one can see that $d$ modifies the speed of the migratory process but it has little influence in the value of equilibrium of urban share of the system, what explains the similar structures seen in all sets of Fig. \ref{JkL}.

\section{Conclusion}
 In this paper we analyzed the rural-urban migration process by means of an agent-based computational model. We extend the analysis carried out in our previous work to lattices with dimensions up to seven.

\par The variation of the dimension of the lattices slightly modify some of the results found in the study of a two-dimensional lattice. To explain this, it is important to remember that the distribution of workers in the lattice does not mean a spatial distribution but the definition of the neighborhood of each individual. Therefore, the bigger the lattice dimension the bigger the neighborhood that each worker will have which is the cause of the small difference in the results.

\section*{Acknowledgments}

\par  Aquino L. Esp\'{\i}ndola thanks CAPES for the financial support. T. J. P. Penna thanks CNPq for the fellowship and Jaylson J. Silveira acknowledges research grants from CNPq.

\end{document}